\documentclass{article}
\usepackage{spconf,amsmath,amssymb,graphicx,hyperref, svg, cite, xurl, multirow}

\title{Segmental Posterior Decoding for Audio Moment Retrieval}
\name{Seungdeok Choi, Seongmin Choi, Inhan Choi, Junho Kim, Jeong-gyu Ban, Yong-Hwa Park\thanks{The code will be made publicly available on GitHub upon acceptance.}}
\address{Department of Mechanical Engineering, Korea Advanced Institute of Science and Technology, Korea}
\begin{document}
%
\maketitle
\begin{abstract}
Audio moment retrieval (AMR) identifies temporal segments in long recordings that best match a free-form text query. Existing systems largely rely on fixed-slot DETR decoders that assign proposal-level confidence scores without explicitly normalizing over competing explanations of the full timeline. We propose segmental posterior decoding, which defines a globally normalized distribution over temporal segmentations and scores each candidate moment by its exact segment marginal posterior computed through forward-backward inference. We further expand the training segmentation space by treating a foreground span and its adjacent subdivisions as distinct hypotheses, thereby increasing competition among alternative segmentations. On CASTELLA, our method achieves 41.15\% R1@0.7 and 34.68\% mAP, outperforming the same network decoded with DETR slot confidence by 10.91 and 9.20 percentage points, respectively.

\end{abstract}
\section{Introduction}
\label{sec:intro}

Audio moment retrieval (AMR) retrieves temporal moments in long, untrimmed audio that correspond to a free-form natural-language query
\cite{munakata2024amr}. Existing systems largely adopt DEtection TRansformer (DETR)-based architectures from video moment retrieval \cite{detr,moon2023query}. Given audio and a text query, an encoder produces query-conditioned audio representations, and a DETR decoder predicts a fixed set of temporal moments associated with confidence scores.

Recent AMR systems have emphasized span localization quality as an important factor in retrieval ranking \cite{sugawara2026_t6,kibata2026_t6, usui2026_t6}. To better align confidence with temporal IoU, they employ Quality Focal Loss (QFL) \cite{ligqfl} and Varifocal Loss (VFL) \cite{Zhang2020VarifocalNetAI}. These approaches improve the confidence assigned to each predicted proposal, but the resulting scores are still estimated at the proposal level rather than derived from a globally normalized distribution over alternative temporal explanations. Consequently, the resulting score does not explicitly quantify the relative plausibility of a candidate moment among competing explanations of the audio.

To model this relative plausibility directly, we introduce segmental posterior decoding. We define a query-conditioned segmental model based on semi-Markov conditional random fields \cite{sarawagi2004semi} and compute the marginal posterior of every candidate moment through forward-backward inference over the entire segmentation space \cite{sarawagi2004semi,lafferty2001conditional}. Instead of ranking candidates by local segment potentials or selecting a single MAP segmentation through Viterbi decoding \cite{viterbi}, we use the segment marginal as the retrieval score. For a candidate moment, the marginal posterior collects probability mass from every possible segmentation in which that moment appears, while normalization over all segmentations makes the score reflect how strongly it is preferred over alternative explanations of the timeline. We further train the segmental model by normalizing over an expanded segmentation space that treats a span and its adjacent subdivisions as distinct segmentation hypotheses, placing the ground-truth segmentation in competition with a richer set of alternative temporal explanations.

\begin{figure*}[!t]
    \centering
    \includegraphics[width=0.85\textwidth,
        trim=0 8.6cm 5cm 2cm,
        clip]{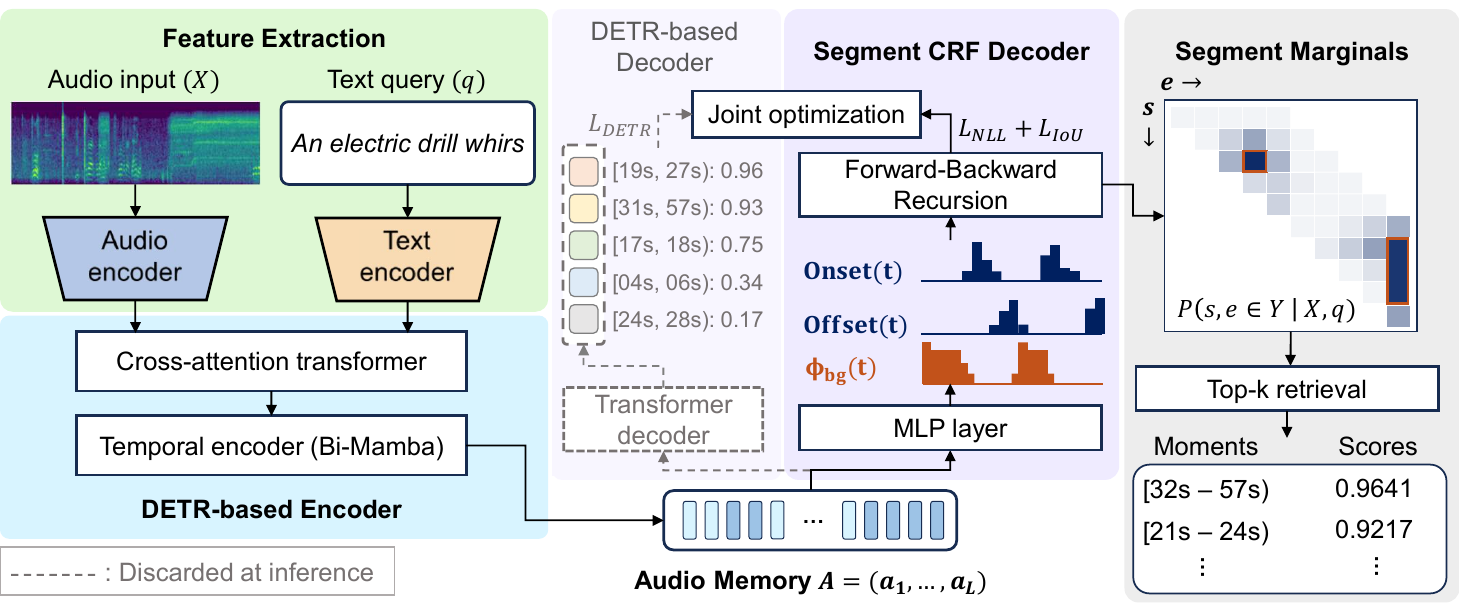}
    \caption{Overview of the proposed segmental posterior decoding framework. Dashed components are used only during training.}
    \label{fig:overview}
\end{figure*}

Experiments on CASTELLA \cite{munakata2026castella} show that segment marginal decoding consistently outperforms proposal-wise confidence scoring on the same network, highlighting the importance of the retrieval decision rule in AMR.\textbf{}

\section{Proposed Method}
\label{sec:method}

To replace proposal-level confidence scoring with a structured retrieval rule, we introduce a segmental CRF decoder that operates on the query-conditioned audio representation of the underlying DETR-based AMR model.
As shown in Fig.~\ref{fig:overview}, the conventional DETR branch trained with Hungarian matching \cite{munakata2024amr, detr} and the proposed segmental decoder share the same audio memory and are jointly optimized during training.
At inference, the DETR decoder is discarded, and candidate moments are ranked by segment marginal posteriors from the segmental decoder, followed by temporal non-maximum suppression (NMS) \cite{NMS}.

\subsection{Segmental CRF Formulation}

We formulate the audio timeline using a segmental conditional random
field (CRF)~\cite{sarawagi2004semi,lafferty2001conditional}.
Foreground (FG) segments represent query-relevant moments, while
uncovered frames are assigned to the background (BG). Given the
query-conditioned audio memory
$\mathbf{A}=(\mathbf{a}_t)_{t=1}^{L}$, a lightweight frame-level head
produces unnormalized onset, offset, and BG scores
$o_t$, $r_t$, and $b_t$, respectively:
\begin{equation}
(o_t,r_t,b_t)=f_{\mathrm{frame}}(\mathrm{LN}(\mathbf{a}_t)).
\label{eq:frame_head}
\end{equation}
For an FG segment $[s,e)$ and BG frame $t$, the segment potential
$\psi$ and BG emission potential $\phi_{\mathrm{bg}}$ are
\begin{equation}
\psi(s,e) = o_s + r_{e-1} + b_{\mathrm{fg}},
\qquad
\phi_{\mathrm{bg}}(t) = b_t,
\label{eq:potential}
\end{equation}
where $o_s$ provides start-boundary evidence at frame $s$,
$r_{e-1}$ provides end-boundary evidence at the last FG frame
$e-1$, and $b_{\mathrm{fg}}$ is a learned segment bias.
Accordingly, $\psi(s,e)$ is an unnormalized log-potential measuring
the compatibility of assigning $[s,e)$ to FG, while
$\phi_{\mathrm{bg}}(t)$ analogously measures the compatibility of
assigning frame $t$ to BG.
Although $\psi(s,e)$ contains no explicit span-interior term, it
remains query-conditioned because the audio memory has already been
conditioned on the text query by the preceding cross-attention encoder.

During training, we additionally allow adjacent FG segments in the
segmentation space. Thus, $\mathrm{FG}[s,e)$ and
$\{\mathrm{FG}[s,m),\mathrm{FG}[m,e)\}$ are distinct hypotheses,
increasing the alternatives that compete with the ground-truth
segmentation under global normalization.

\begin{table*}[t]
\centering
\caption{
Comparison with existing AMR systems on CASTELLA and UnAV-100.
CASTELLA results are reported on the dev-test split, while UnAV-100 is evaluated
zero-shot using the same CASTELLA-trained checkpoints without additional training.
All results are mean $\pm$ standard deviation over three seeds.
Best results are shown in bold.
}
\label{tab:main_results}
\resizebox{\textwidth}{!}{
\begin{tabular}{lccc|ccc}
\hline
\multirow{2}{*}{Method}
& \multicolumn{3}{c|}{CASTELLA}
& \multicolumn{3}{c}{UnAV-100 (zero-shot)} \\
& R1@0.5 & R1@0.7 & mAP
& R1@0.5 & R1@0.7 & mAP \\
\hline

DETR architectural baseline~\cite{choi2026task6}
& $47.41{\pm}1.09$ & $30.76{\pm}0.47$ & $25.28{\pm}0.62$
& $63.00{\pm}2.65$ & $51.33{\pm}3.51$ & $49.22{\pm}1.43$ \\

\quad + QFL~\cite{ligqfl}
& $50.04{\pm}0.46$ & $32.96{\pm}0.41$ & $26.23{\pm}0.73$
& $64.33{\pm}3.06$ & $48.00{\pm}4.36$ & $46.95{\pm}4.17$ \\

\quad + VFL~\cite{Zhang2020VarifocalNetAI}
& $51.05{\pm}1.15$ & $33.14{\pm}0.45$ & $26.52{\pm}0.68$
& $63.00{\pm}1.73$ & $47.33{\pm}1.53$ & $46.21{\pm}0.93$ \\

\textbf{Segmental posterior decoding (ours)}
& $\mathbf{53.95{\pm}0.49}$ & $\mathbf{41.15{\pm}0.24}$ & $\mathbf{34.68{\pm}0.17}$
& $\mathbf{72.33{\pm}2.08}$ & $\mathbf{61.00{\pm}2.65}$ & $\mathbf{60.33{\pm}1.60}$ \\
\hline
\end{tabular}
}
\end{table*}

\subsection{Segment Marginal Posterior Decoding}
\label{sec:marginal_decoding}

Using the local potentials above, each segmentation $Y$ is scored as
\begin{equation}
S(Y)=
\sum_{(s,e)\in Y}\psi(s,e)
+
\sum_{t\in \mathrm{BG}(Y)}\phi_{\mathrm{bg}}(t).
\label{eq:seg_score}
\end{equation}
Although $\psi(s,e)$ depends only on the boundaries, frames inside an FG segment no longer contribute their BG potentials $\phi_{\mathrm{bg}}$, so interior evidence still affects the score of the complete segmentation. We define the globally normalized segmentation
posterior as

\begin{equation}
P(Y\mid X,q)=\frac{\exp S(Y)}{Z(X,q)},\qquad
Z(X,q)=\sum_{\tilde{Y}\in\mathcal{Z}}\exp S(\tilde{Y}),
\label{eq:seg_posterior}
\end{equation}
where $\mathcal{Z}$ denotes the segmentation space.

The partition function and segment marginals are computed exactly by
semi-Markov forward-backward inference~\cite{sarawagi2004semi}.
Let $\alpha_t$ and $\beta_t$ aggregate all valid segmentations of
$[0,t)$ and $[t,L)$, respectively, and
$D_t=\min(t,D_{\max})$. The forward recursion is
\begin{equation}
\begin{aligned}
\alpha_t = \operatorname{LSE}\Big(
&\alpha_{t-1}+\phi_{\mathrm{bg}}(t-1),\\
&\operatorname{LSE}_{1\le d\le D_t}
  [\alpha_{t-d}+\psi(t-d,t)]
\Big),
\end{aligned}
\label{eq:forward}
\end{equation}
where \(\operatorname{LSE}\) denotes the log-sum-exp operator with $\alpha_0=0$, while $\beta_t$ is computed analogously and
$\log Z=\alpha_L$. The exact segment marginal posterior is
\begin{equation}
P((s,e)\in Y\mid X,q)
=
\exp\!\left(
\alpha_s+\psi(s,e)+\beta_e-\log Z
\right),
\label{eq:marginal}
\end{equation}
which is used directly as the retrieval score.

Because mutually exclusive overlapping candidates can each receive high marginal probability by being supported by different segmentations, we apply greedy temporal NMS to
the top-$K_{\mathrm{pool}}$ candidates and retain the top-$K$
non-redundant moments.

\subsection{Training Objective}
\label{sec:objective}

We optimize the segmental decoder with two complementary objectives: a negative
log-likelihood (NLL) objective and a minimum-IoU-risk objective.
First, we minimize the negative log-likelihood of the ground-truth segmentation
$Y^\ast$:
\begin{equation}
    \mathcal{L}_{\mathrm{NLL}}
    =
    \frac{1}{L}
    \left(
    \log Z-S(Y^\ast)
    \right),
\end{equation}
where the loss is normalized by the valid sequence length $L$.

To further align the segment marginals with the temporal IoU used for AMR
evaluation, we employ a minimum-risk objective \cite{ZHOU20141904}.
Let $G$ denote the set of ground-truth moments and
$\rho(s,e)=\max_{g\in G}\operatorname{IoU}([s,e),g)$ denote the best overlap
between candidate $[s,e)$ and any ground-truth moment.
Using the exact segment marginals from Sec.~\ref{sec:marginal_decoding}, we
compute
\begin{equation}
    \mathcal{L}_{\mathrm{IoU}}
    =
    1-
    \frac{
        \sum_{s,e}
        P((s,e)\in Y\mid X,q)\,\rho(s,e)
    }{
        \sum_{s,e}
        P((s,e)\in Y\mid X,q)
    } .
\end{equation}

The two objectives serve different roles: the NLL favors the ground-truth
segmentation under global normalization, whereas the IoU-risk encourages
higher segment marginals for temporally accurate candidates.
The complete training objective is
$\mathcal{L}_{\mathrm{DETR}}
+\lambda_{\mathrm{NLL}}\mathcal{L}_{\mathrm{NLL}}
+\lambda_{\mathrm{risk}}\mathcal{L}_{\mathrm{risk}}$, where $\mathcal{L}_{\mathrm{DETR}}$ denotes the original DETR training loss.

\section{Experimental Settings}
\label{sec:exp}

We follow the two-stage DCASE 2026 Task~6 training protocol~\cite{dcase2026task6},
pre-training on Clotho-Moment~\cite{munakata2024amr} and fine-tuning on
CASTELLA~\cite{munakata2026castella}. Evaluation is conducted on the
CASTELLA dev-test split with 1,347 queries and on UnAV-100~\cite{geng2023dense}
for zero-shot transfer. We use the single-encoder, single-decoder DETR architecture of~\cite{choi2026task6} with M2D-CLAP features~\cite{niizumi2025m2d-clap}.

Both stages use AdamW with batch size 32 for up to 200 epochs, with
learning rates of $1e{-4}$ and $8e{-5}$ for pre-training
and fine-tuning, respectively. Marginal decoding draws the top 50
candidates, applies temporal NMS at IoU 0.7, and retains the top 10.
All main results are averaged over three seeds and reported as
mean $\pm$ standard deviation. Evaluation follows the DCASE 2026 Task 6 protocol using R1@0.5 and R1@0.7,
where R1 denotes Recall@1 and the suffix specifies the temporal IoU threshold,
together with mean Average Precision (mAP) averaged over IoU thresholds from
0.5 to 0.95.

\begin{table*}[t]
\centering
\caption{
Analysis of segmental decoding on CASTELLA.
(a) compares alternative decoding rules using the same trained checkpoints. 
(b) evaluates additions to the minimal segment potential, while
(c) removes components of the full training design.
Rows in (b) and (c) are modifications of the Segment marginal configuration in (a).
Short and Long denote R1@0.7 for moments shorter than 3\,s and at least 10\,s, respectively.
Results are mean $\pm$ standard deviation over three seeds.
}
\label{tab:decoding_analysis}
\small
\setlength{\tabcolsep}{8pt}
\begin{tabular}{lccccc}
\hline
& R1@0.5 & R1@0.7 & mAP & Short $(<3s)$ & Long $(\geq10s)$ \\
\hline

\multicolumn{6}{l}{\textit{(a) Decoding rule}} \\

DETR decoder
& $47.81{\pm}0.45$
& $30.24{\pm}0.69$
& $25.48{\pm}0.31$
& $12.15{\pm}1.04$
& $45.98{\pm}1.18$ \\

Local potential $\psi(s,e)$
& $39.92{\pm}1.11$
& $29.45{\pm}0.48$
& $25.78{\pm}0.32$
& $22.19{\pm}0.47$
& $37.13{\pm}0.26$ \\

Viterbi MAP
& $41.94{\pm}1.67$
& $33.28{\pm}1.19$
& $22.29{\pm}0.99$
& $17.90{\pm}2.20$
& $48.58{\pm}0.93$ \\

\textbf{Segment marginal}
& $\mathbf{53.95{\pm}0.49}$
& $\mathbf{41.15{\pm}0.24}$
& $\mathbf{34.68{\pm}0.17}$
& $26.58{\pm}1.10$
& $\mathbf{52.53{\pm}1.27}$ \\

\hline
\multicolumn{6}{l}{
\textit{(b) Additions to }
$\psi_{\mathrm{base}}(s,e)$ \textit{from eq.~\ref{eq:potential} }
} \\

$+\;$ span-mean content
& $53.95{\pm}1.06$
& $41.03{\pm}0.95$
& $34.55{\pm}0.12$
& $\mathbf{28.22{\pm}2.39}$
& $51.56{\pm}0.23$ \\

$+\;$ query-conditioned duration prior
& $51.08{\pm}0.84$
& $36.90{\pm}1.12$
& $33.32{\pm}0.43$
& $25.21{\pm}1.37$
& $46.65{\pm}2.20$ \\

$+\;$ shared duration prior
& $51.65{\pm}0.73$
& $37.32{\pm}0.60$
& $33.12{\pm}0.71$
& $24.75{\pm}1.24$
& $46.88{\pm}0.97$ \\

\hline
\multicolumn{6}{l}{\textit{(c) Training design}} \\

$-\;$ IoU-risk
& $47.51{\pm}0.90$
& $34.47{\pm}1.18$
& $32.16{\pm}0.41$
& $26.30{\pm}1.71$
& $41.30{\pm}1.02$ \\

$-\;$ Adjacent FG segments
& $51.32{\pm}0.45$
& $37.76{\pm}0.50$
& $33.52{\pm}0.34$
& $24.38{\pm}0.28$
& $47.40{\pm}1.10$ \\

\hline
\end{tabular}
\end{table*}

\section{Results \& Discussions}

\subsection{Overall Performance and Cross-Dataset Transfer}

Table~\ref{tab:main_results} first compares segmental posterior decoding with the DETR-based baseline~\cite{choi2026task6}. Compared with the DETR architectural baseline, segmental posterior decoding substantially improves CASTELLA performance, particularly at the stricter IoU threshold and in mAP, indicating that the retrieval decision rule remains a major bottleneck.

We next compare with QFL and VFL, which explicitly incorporate localization quality into proposal confidence. Although both improve over conventional DETR scoring, segmental posterior decoding remains substantially stronger, outperforming VFL by 8.02 points in R1@0.7 and 8.16 points in mAP. This shows that the gain extends beyond improved proposal-wise localization calibration and benefits from globally normalized competition among alternative temporal explanations.

Segmental posterior decoding achieves 41.15 R1@0.7 on CASTELLA dev-testing, compared with 41.13 \cite{ogawa2026_t6} and 40.70 \cite{kibata2026_t6} for the strongest non-ensemble, non-LLM submissions on the DCASE 2026 Task~6 leaderboard \cite{dcase2026task6results}. This places the proposed method at the level of the strongest single-model systems and within several points of the overall best result of 45.14 \cite{kim2026_t6}, which additionally employs model ensembling and LLM-based refinement.

The same trend largely transfers to UnAV-100 without additional training. QFL and VFL no longer improve over the DETR baseline, whereas segmental posterior decoding retains a clear advantage across all metrics, indicating that its benefit is not confined to the CASTELLA domain.

\subsection{Ablation on Segmental Decoding}

Table~\ref{tab:decoding_analysis}(a) isolates the effect of the decoding rule using
the same trained checkpoints. Segment marginal decoding outperforms DETR slot confidence, local-potential ranking, and Viterbi
decoding in both R1 and mAP. The alternatives fail differently:
DETR confidence is particularly weak for short moments, whereas the
local potential degrades most on long moments. Unlike the local
potential or a single Viterbi segmentation, segment marginals score each
candidate by aggregating its probability over all compatible
segmentations, accounting for competing explanations of the timeline.

Table~\ref{tab:decoding_analysis}(b) shows that this gain does not
require a more expressive segment potential. Adding span-mean
audio-text content provides no meaningful improvement despite nearly
tripling the active parameters in the segmental head
(67k$\rightarrow$199k), likely because the segment is already
text-conditioned. Adding either duration prior further degrades
localization, particularly for long moments. These results indicate
that the improvement primarily arises from structured marginalization
rather than a complex local segment scorer.

\subsection{Training Objective and Expanded Segmentation Space}

Table~\ref{tab:decoding_analysis}(c) analyzes the training design.
Removing the IoU-risk substantially degrades localization, particularly
for long moments, indicating that the metric-aligned risk objective is
important for accurate temporal localization.
Disallowing adjacent FG segments also degrades both R1 and mAP.
Despite similar training likelihoods, allowing adjacent FG segments lowers
posterior entropy (2.02$\rightarrow$1.79), and increases the marginal mass assigned to ground-truth and peak segments consistently across all three seeds.
These results suggest that allowing adjacent FG segments during training
strengthens competition among alternative segmentations under global
normalization, and encourages a more concentrated posterior.

\begin{figure}[t]
    \centering
    \includegraphics[width=0.90\linewidth,
        trim=0 0cm 0cm 0cm,
        clip]{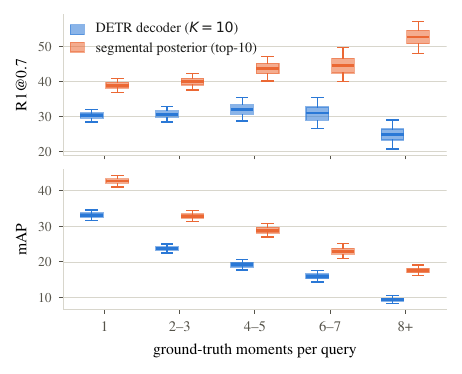}
    \caption{R1@0.7 and mAP stratified by the number of ground-truth moments per query.}
    \label{fig:count_stratified}
\end{figure}

\subsection{Effect of the Number of Moments}

Fig.~\ref{fig:count_stratified} shows that the advantage of segment
marginal decoding grows with the number of ground-truth moments.
While DETR performance degrades in the high-count regime, segment
marginal decoding remains robust and achieves an increasingly larger
R1@0.7 advantage. mAP decreases for both methods as the number of ground-truth moments
increases, reflecting the greater difficulty of recovering a larger set
of relevant moments with both methods limited to ten retained predictions. Nevertheless,
segment marginal decoding maintains a clear advantage across all
moment-count groups. 

\section{Conclusion}

We presented segmental posterior decoding for AMR, replacing
proposal-level confidence with exact segment marginal posteriors under
a globally normalized segmental model. Experiments show that the main gain arises from structured marginal decoding, whose advantage increases for queries with multiple relevant moments. Competitive normalization over the expanded segmentation space provides an additional improvement. Future
work will explore audio-specific segment potentials that explicitly
model acoustic structure within candidate moments.

\vfill\pagebreak

\section{Acknowledgments}
This work was supported by the Korea Institute for Advancement of Technology (KIAT) grant funded by the Korea Government (MOTIE) (RS-2025-02263945, HRD Program for Industrial Innovation).

\section{Compliance with Ethical Standards}
This research used publicly available datasets and involved no direct experimentation on human or animal subjects. Therefore, no additional ethical approval was required for this study.

\let\oldthebibliography\thebibliography
\renewcommand{\thebibliography}[1]{%
  \oldthebibliography{#1}%
  \setlength{\itemsep}{0pt}%
  \setlength{\parskip}{0pt}%
  \setlength{\parsep}{0pt}%
}

\bibliographystyle{IEEEbib}
\bibliography{strings,refs}

@inproceedings{sarawagi2004semi,
 author = {Sarawagi, Sunita and Cohen, William W},
 booktitle = {Advances in Neural Information Processing Systems},
 pages = {},
 title = {Semi-Markov Conditional Random Fields for Information Extraction},
 url = {https://proceedings.neurips.cc/paper_files/paper/2004/file/eb06b9db06012a7a4179b8f3cb5384d3-Paper.pdf},
 volume = {17},
 year = {2004}
}

@inproceedings{lafferty2001conditional,
  title     = {{Conditional Random Fields: Probabilistic Models for Segmenting and Labeling Sequence Data}},
  author    = {Lafferty, John D. and McCallum, Andrew and Pereira, Fernando C. N.},
  booktitle = {Int. Conf. Mach. Learn. (ICML)},
  year      = {2001},
  pages     = {282-289},
  url       = {https://mlanthology.org/icml/2001/lafferty2001icml-conditional/}
}

@inproceedings{munakata2024amr,
    author = "Munakata, Hokuto and Nishimura, Taichi and Nakada, Shota and Komatsu, Tatsuya",
    title = "Language-based Audio Moment Retrieval",
    booktitle = "Proc. IEEE Int. Conf. Acoust., Speech and Signal Process. (ICASSP)",
    year = "2025",
    pages={1-5},
}

@inproceedings{munakata2026castella,
    author = "Munakata, Hokuto and Imamura, Takehiro and Nishimura, Taichi and Komatsu, Tatsuya",
    title = "{CASTELLA}: Long Audio Dataset with Captions and Temporal Boundaries",
    booktitle = "Proc. IEEE Int. Conf. Acoust., Speech and Signal Process. (ICASSP)",
    year = "2026",
    pages={15352-15356},
}

@inproceedings{detr,
author = {Carion, Nicolas and Massa, Francisco and Synnaeve, Gabriel and Usunier, Nicolas and Kirillov, Alexander and Zagoruyko, Sergey},
title = {End-to-End Object Detection with Transformers},
year = {2020},
isbn = {978-3-030-58451-1},
url = {https://doi.org/10.1007/978-3-030-58452-8_13},
doi = {10.1007/978-3-030-58452-8_13},
booktitle = {Proc. Eur. Conf. Comput. Vis. (ECCV)},
pages = {213–229},
numpages = {17},
location = {Glasgow, United Kingdom}
}

@inproceedings{moon2023query,
  title={Query-dependent video representation for moment retrieval and highlight detection},
  author={Moon, WonJun and Hyun, Sangeek and Park, SangUk and Park, Dongchan and Heo, Jae-Pil},
  booktitle={Proc. IEEE/CVF Conf. Comput. Vis. Pattern Recognit. (CVPR)},
  pages={23023--23033},
  year={2023}
}

@techreport{sugawara2026_t6,
    Author = "Sugawara, Haruto and Nakamura, Tomoko and Sakaguchi, Ken and Aida, Hikari and Yuri, Shunta",
    title = "{YCU} SUBMISSION FOR {DCASE} 2026 CHALLENGE TASK 6",
    institution = "DCASE2026 Challenge",
    year = "2026",
    month = "June",
}

@techreport{usui2026_t6,
    Author = "Usui, Ren and Fujimoto, Ryuta and Kikuchi, Mikuri and Kitao, Taichi and Suzuki, Tomohisa",
    title = "IMPROVING TEMPORAL BOUNDARY PRECISION IN AUDIO MOMENT RETRIEVAL",
    institution = "DCASE2026 Challenge",
    year = "2026",
    month = "June",
}

@ARTICLE{viterbi,
  author={Viterbi, A.},
  journal={IEEE Transactions on Information Theory}, 
  title={Error bounds for convolutional codes and an asymptotically optimum decoding algorithm}, 
  year={1967},
  volume={13},
  number={2},
  pages={260-269},
  doi={10.1109/TIT.1967.1054010}}

@INPROCEEDINGS{ligqfl,
  author={Li, Xiang and Wang, Wenhai and Hu, Xiaolin and Li, Jun and Tang, Jinhui and Yang, Jian},
  booktitle={Proc. IEEE/CVF Conf. Comput. Vis. Pattern Recognit. (CVPR)}, 
  title={Generalized Focal Loss V2: Learning Reliable Localization Quality Estimation for Dense Object Detection}, 
  year={2021},
  volume={},
  number={},
  pages={11627-11636},
  doi={10.1109/CVPR46437.2021.01146}}

@article{Zhang2020VarifocalNetAI,
  title={VarifocalNet: An IoU-aware Dense Object Detector},
  author={Haoyang Zhang and Ying Wang and Feras Dayoub and Niko Sunderhauf},
  journal={Proc. IEEE/CVF Conf. Comput. Vis. Pattern Recognit. (CVPR)},
  year={2021},
  pages={8510-8519},
  url={https://api.semanticscholar.org/CorpusID:221376816}
}

@INPROCEEDINGS{NMS,
  author={Neubeck, A. and Van Gool, L.},
  booktitle={Int. Conf. on Pattern Recognit. (ICPR)}, 
  title={Efficient Non-Maximum Suppression}, 
  year={2006},
  volume={3},
  number={},
  pages={850-855},
  doi={10.1109/ICPR.2006.479}}

@article{ZHOU20141904,
title = {Minimum-risk training for semi-Markov conditional random fields with application to handwritten Chinese/Japanese text recognition},
journal = {Pattern Recognition},
volume = {47},
number = {5},
pages = {1904-1916},
year = {2014},
issn = {0031-3203},
doi = {https://doi.org/10.1016/j.patcog.2013.12.002},
url = {https://www.sciencedirect.com/science/article/pii/S0031320313005207},
author = {Xiang-Dong Zhou and Yan-Ming Zhang and Feng Tian and Hong-An Wang and Cheng-Lin Liu},
}

@techreport{choi2026task6,
    Author = "Choi, Seungdeok and Yong-Hwa Park",
    title = "MULTI-SIGNAL CASCADED GROUNDING FOR AUDIO MOMENT RETRIEVAL FROM LONG AUDIO",
    institution = "DCASE2026 Challenge",
    year = "2026",
    month = "June",
}

@misc{dcase2026task6,
  author       = {{DCASE Community}},
  title        = {{DCASE} 2026 Challenge Task 6: Audio Moment Retrieval from Long Audio},
  year         = {2026},
  howpublished = {[Online]. Available: \url{https://dcase.community/challenge2026/task-audio-moment-retrieval-from-long-audio}},
  note         = {Accessed: Sep. 12, 2026}
}

@article{niizumi2025m2d-clap,
    author  = {Niizumi, Daisuke and Takeuchi, Daiki and Yasuda, Masahiro and Nguyen, Binh Thien and Ohishi, Yasunori and Harada, Noboru},
    journal = {IEEE Access}, 
    title   = {{M2D-CLAP: Exploring General-purpose Audio-Language Representations Beyond CLAP}}, 
    year    = {2025},
    volume  = {13},
    pages   = {163313-163330},
    doi={10.1109/ACCESS.2025.3611348}}

@techreport{kibata2026_t6,
    Author = "Kibata, Koki and Yamamoto, Sayaka and Kawabata, Tomoki and Higashino, Yuma and Osawa, Yuki",
    title = "ADVANCED AUDIO MOMENT RETRIEVAL VIA {CG-DETR}",
    institution = "DCASE2026 Challenge",
    year = "2026",
    month = "June",
}

@misc{dcase2026task6results,
  author       = {{DCASE Community}},
  title        = {{DCASE} 2026 Challenge Task 6: Audio Moment Retrieval from Long Audio -- Results},
  year         = {2026},
  howpublished = {[Online]. Available: \url{https://dcase.community/challenge2026/task-audio-moment-retrieval-from-long-audio-results}},
  note         = {Accessed: Sep. 12, 2026}
}

@inproceedings{geng2023dense,
    title={Dense-Localizing Audio-Visual Events in Untrimmed Videos: A Large-Scale Benchmark and Baseline},
    author={Geng, Tiantian and Wang, Teng and Duan, Jinming and Cong, Runmin and Zheng, Feng},
    booktitle={Proc. IEEE/CVF Conf. Comput. Vis. Pattern Recognit.},
    pages={22942--22951},
    year={2023}
    }

@techreport{ogawa2026_t6,
    Author = "Ogawa, Takumi and Ohhashi, Nobuyuki and Kurisuno, Yota and Takenaka, Minami and Miyamoto, Ririko",
    title = "ENHANCED AUDIO MOMENT RETRIEVAL APPROACH FOR {DCASE} 2026 TASK 6",
    institution = "DCASE2026 Challenge",
    year = "2026",
    month = "June",
}

@techreport{kim2026_t6,
    Author = "Kim, JeongRae and Jung, ho jun and Park, Yewon and Lim, Changwon",
    title = "{QAM-DETR} SYSTEM FOR {DCASE} 2026 TASK 6: QUALITY-AWARE MAMBA {DETR} FOR QUERY-BASED AUDIO MOMENT RETRIEVAL",
    institution = "DCASE2026 Challenge",
    year = "2026",
    month = "June",
}

\end{document}